\documentclass[aps,prb,twocolumn,superscriptaddress,preprintnumbers]{revtex4-2}
\usepackage{amsmath}
\usepackage[version=4]{mhchem}
\usepackage{graphicx} 
\usepackage{caption}
\usepackage{subcaption}
\usepackage{ragged2e}
\usepackage{xcolor}
\usepackage{geometry}
\usepackage{hyperref}
\usepackage[normalem]{ulem}

\hypersetup{
    colorlinks=true,
    linkcolor=blue,
    filecolor=magenta,      
    urlcolor=blue,
    citecolor=blue,
    }

\newcommand{\hso}{\ce{Ho3ScO6}}

\begin{document}

\title{Structure and Magnetic Properties of a Maple Leaf antiferromagnet Ho$_3$ScO$_6$}

\author{C. Aguilar-Maldonado}
\affiliation{Helmholtz-Zentrum Berlin f\"{u}r Materialien und Energie GmbH, Hahn-Meitner Platz 1, D-14109 Berlin, Germany}

\author{R.~Feyerherm}
\affiliation{Helmholtz-Zentrum Berlin f\"{u}r Materialien und Energie GmbH, Hahn-Meitner Platz 1, D-14109 Berlin, Germany}

\author{K.~Proke\v{s}}
\affiliation{Helmholtz-Zentrum Berlin f\"{u}r Materialien und Energie GmbH, Hahn-Meitner Platz 1, D-14109 Berlin, Germany}

\author{L.~Keller}
\affiliation{Laboratory for Neutron Scattering and Imaging, Paul Scherrer Institut, Villigen CH-5232, Switzerland}

\author{B.~Lake}
\affiliation{Helmholtz-Zentrum Berlin f\"{u}r Materialien und Energie GmbH, Hahn-Meitner Platz 1, D-14109 Berlin, Germany}
\affiliation{Institut f\"{u}r Festk\"{o}rperphysik, Technische Universit\"{a}t Berlin, Hardenbergstra{\ss}e 36, D-10623 Berlin, Germany}

\begin{abstract}
\hso~harbours a frustrated Maple Leaf Lattice (MLL). It crystalizes in the \ce{Mg3TeO6}-type structure, and has a centrosymmetric trigonal space group (R$\bar{3}$). This system contains stacked layers of magnetic rings along the c-axis consisting of six magnetic \ce{Ho^3+}ions forming Ho hexagons, which are connected into a 2-dimensional network by equilateral and isosceles triangles to form a rare example of a MLL. Long range magnetic order is reached below $T_N=4.1$~K with a 120$^\circ$ spin arrangement on the equilateral triangles resulting in a positive vector chirality ground state configuration. 
 
\end{abstract}

\maketitle

\section{Introduction}

Magnetic frustration has gained a lot of interest in the area of solid state physics as it has been the basis for the discovery of new exotic physics. Compounds with localized magnetic moments that show absence of long-range magnetic ordering at low temperatures despite strong magnetic interactions offer interesting new physical phenomena and degenerate ground states \cite{KHATUA20231}. Geometric frustration in spin systems can be found in a variety of crystallographic structures where the magnetic ions form specific lattices which are typically built from triangular or tetrahedral units in which the ions interact antiferromagnetically. Examples of two-dimensional (2D) frustrated magnets are found on triangular \cite{Li_2020,PhysRevB.110.054442,PhysRevB.95.144404} and Kagome lattices \cite{doi:10.1126/science.aaw1666,MENDELS2016455,Balz2016,Shi2022}, meanwhile a well-know example of a three-dimensional (3D) frustrated lattice is the pyrochlore lattice \cite{OGUNBUNMI2021100552,Benton2016,PhysRevB.92.224430}.

In the triangular lattice, each lattice point has $z=6$ neighbors. In contrast, in the Kagome lattice which can be achieved by a 1/4 depletion of lattice sites from the triangular lattice \cite{Aliev2012}, $z$ is reduced to 4. The Maple Leaf Lattice (MLL) is a much less explored type of 2D lattice which is obtained by depleting 1/7 of the sites from the triangular lattice to give $z=5$ \cite{PhysRevB.65.224405,ActaPolA972000}. It can also be geometrically frustrated in the presence of antiferromagnetic interactions between the magnetic ions on the lattice sites. A characteristic of the MLL lattice is the existence of three nonequivalent nearest-neighbor interactions, unlike the Kagome and triangular lattices where there is only one type of nearest-neighbor interaction. These three nearest neighbor interactions are respectively: $J_{d}$ which couples only two lattice sites into a dimer, $J_{t}$ which couples three sites into an equilateral triangle and $J_{h}$ which couples six sites into a hexagon, see Fig.~\ref{fig:str-latt}(c).

\begin{figure}
    \centering
    \includegraphics[width=1\linewidth]{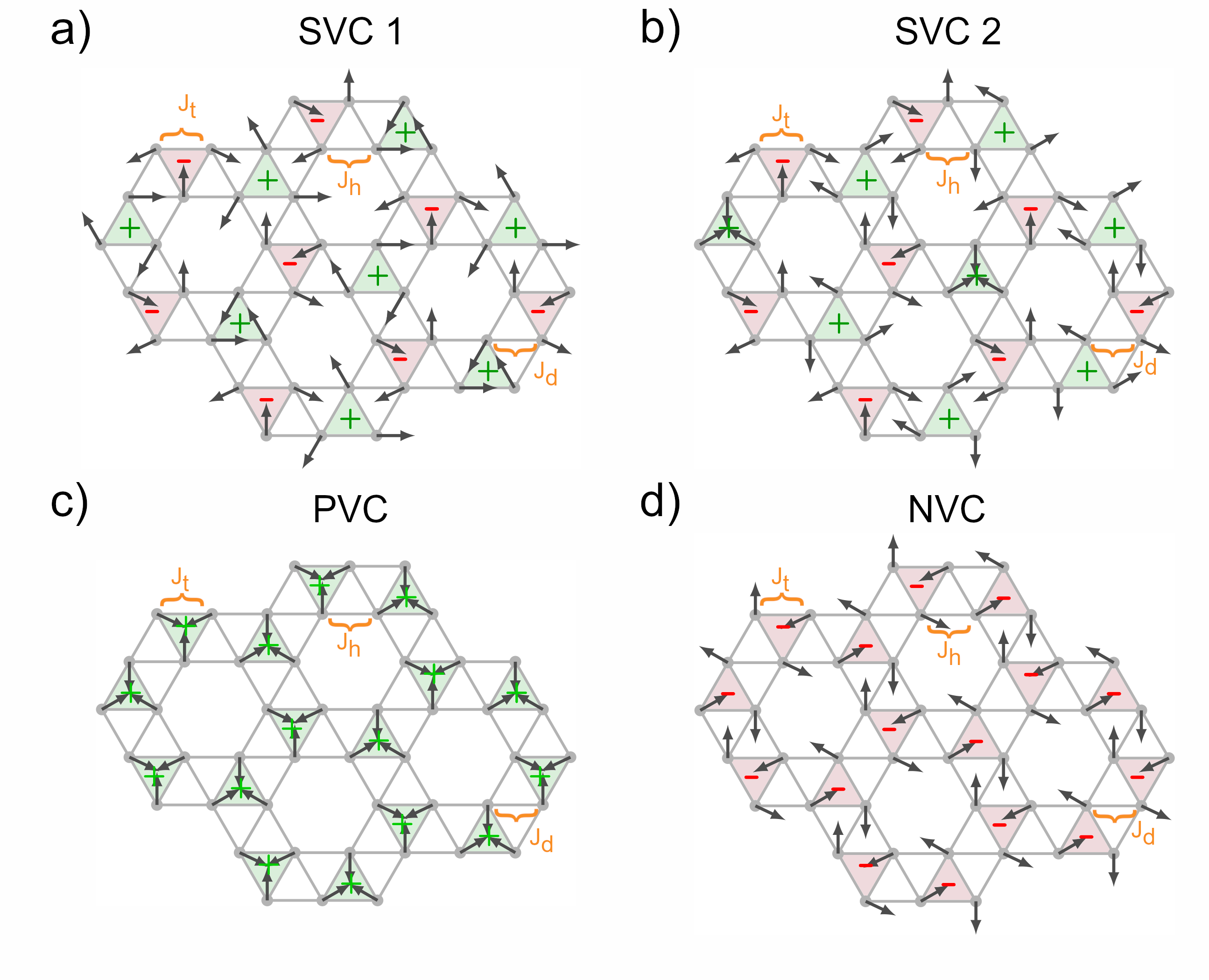}
    \captionsetup{justification=justified, singlelinecheck=false}
    \caption{\justifying Maple Leaf Lattice with threee kind of magnetic interactions ($J_h$, $J_d$, $J_t$). a) The black arrow represent the spin arrangement for the particular case where $J_{d}=J_{t}=J_{h}$ namely staggered vector chirality 1 (SCV 1) and three possible solutions for the case $J_{d} >> J_{t} > J_{h}$: b) staggered vector chirality 2 (SCV 2), c) positive vector chirality (PVC), and negative vector chirality (NVC). The green plus and red negative symbols refer to the up and down vector chiralities, respectively.}
    \label{fig:intro}
\end{figure}

Spin-1/2 maple leaf lattice antiferromagnets are expected to exhibit interesting phenomena due to frustration effects combined with quantum fluctuations such as for Spangolite, \ce{Cu6Al(SO4)(OH)12Cl*3H2O} \cite{Fennell_2011,Hawthorne93} and bluebellite \ce{Cu6IO3(OH)10Cl} \cite{PhysRevB.110.094406,Mills_Kampf_Christy_Housley_Rossman_Reynolds_Marty_2014}. Another well-studied MLL is \ce{MgMn3O7*3H2O} whose bulk magnetic and thermodynamic properties have been investigated showing it to be a Heisenberg antiferromagnet with classical spin-3/2. Special attention has been paid to the study of chiral magnetic order \cite{Grohol2005, PhysRevB.98.064412, PhysRevB.109.184422, PhysRevB.105.L180412} on the MLL concluding that the ratio between the nearest neighbor interactions determines the magnetic structure and type of chirality of such materials.  In the special case of $J_{d}=J_{t}=J_{h}$, vector chirality emerges in a 120° spin order on every equilateral triangle coupled by $J_{t}$, as in the triangular lattice, the sign of this vector alternates giving the staggered vector chiral 1 (SVC1) spin structure Fig \ref{fig:intro}(a) \cite{PhysRevB.98.064412}. When $J_d$ is allowed to be different from other interactions, it was shown that every $J_t$ triangle exhibits 120$^\circ$ order, but with the relative angle between spins on different triangles varying as a function of $J_d$, and for large $J_d$ the spins coupled by $J_d$ are constrained to be antiparallel \cite{PhysRevB.109.184422,PhysRevB.105.L180412}. For $J_{d} >> J_{t} > J_{h}$ three magnetic structures are possible: for staggered vector chirality 2 (SVC2) (Fig \ref{fig:intro}(b)) the $J_{t}$ triangles keep the staggered arrangement of SCV1. In addition a positive vector chirality (PVC) where the chirality for every $J_{t}$ triangle is maintained positive Fig \ref{fig:intro}(c) and the counterpart named negative vector chirality (NVC) (Fig \ref{fig:intro}(d)) where the $J_{t}$ triangles keeps a negative chirality are also possible \cite{PhysRevB.98.064412,PhysRevB.107.064419}.

Overall, maple leaf lattice compounds offer a rich playground for investigating novel magnetic phenomena and understanding the interplay between lattice structures and magnetic properties. MLL compounds can be realized within the double oxides family  which includes double perovskites and double corundum \ce{A2BB{'}O6} which is of great importance because of its potential for strong and unusual magnetic interactions. Due to small tolerance factors, it is difficult to synthesize such materials at ambient pressure and temperature. Some ambient pressure compositions are found in the literature to present interesting magnetic structures \cite{VASALA20151}. An example is \ce{Mn3TeO6} (space group R$\bar{3}$) which consists of buckled MLL layers. However a complex interaction scheme including frustrated magnetic exchange coupling  gives rise to an elliptical helix and sinusoidal spin density wave coexisting below $T=24$~K while below 21~K it shows multiferroic behavior \cite{Pssr.201510347,C9CC07733B}. 

In this paper we study a corundum material which harbours a frustrated MLL, namely \hso. This material was previously reported although its magnetic properties have never been studied before \cite{BADIE1978311, BADIE1973}. It crystallizes in the \ce{Mg3TeO6}-type structure which has a centrosymmetric trigonal space group R$\bar{3}$ and has cell parameters $a=b=9.469$~\AA~ $c=10.912$~\AA. The magnetic \ce{Ho^{3+}} ions occupy one special Wyckoff position (18f) in the unit cell and are arranged in buckled hexagonal rings in the {\bf a}-{\bf b}-plane forming a MLL which are stacked along the c-axis in an ABC-type fashion. This work reports the first study of the magnetic properties and magnetic structure of \hso. Magnetization, heat capacity and neutron diffraction were carried out on a polycrystalline sample. The compound exhibits antiferromagnetic properties with a low frustration parameter, and shows a decrease in the transition temperature with increasing magnetic field strength. The magnetic structure has been elucidated whereby the chirality of each $J_{t}$ triangle is identical and opposite to that of each hexagon $J_{h}$.

\section{Experimental Methods}

Powder samples were prepared by the solid state method at the CoreLab Quantum Materials (CLQM), Helmholtz Zentrum Berlin f\"{u}r Materialien und Energie (HZB), Germany. Stoichiometric amounts of \ce{Sc2O3} (Aldrich 99.999\%) and \ce{Ho2O3} (Aldrich 99.999\%) were ground, mixed, pressed into pellets and heated up to 1300$^{\circ}$C for 12 hours with cooling and heating rates of 3$^{\circ}$C/min. The precursors were previously heated for 12 hour at 900$^{\circ}$C to eliminate moisture. To ensure the single phase character of the sample, powder X-ray diffraction was collected at room temperature on a Bruker D8 diffractometer equipped with a Cu tube (40kV, 30mA) and a monochromator giving a wavelength $\lambda_{K_{\alpha 1}}=1.54059$ \AA. 

Magnetic measurements as a function of temperature and field were collected using a Quantum Design (QD) SQUID magnetometer (MPMS3). Zero field cooled and field-cooled magnetization curves were measured over a temperature range from 2~K to 350~K at H=0.1 T. Magnetization curves as a function of field were collected at different temperatures ranging from $T=2$~K to $T=10$~K over a field range from $H=-14$~T to $H=14$~T. Heat capacity measurements as a function of temperature were performed over a range from 2~K to 300~K and fields ranging from 0~T to 9~T on a QD Physical Property Measurements System (PPMS). 

A neutron powder diffraction (NPD) experiment was performed on a 4~g polycrystalline sample of \hso~ to investigate its magnetic structure using the SINQ cold neutron diffractometer DMC at the Paul Scherrer Institute (PSI), Switzerland with $\lambda=2.453$~\AA. A temperature variation ranging from 1.6 K to 150 K was carried out to follow the evolution of the magnetic structure with temperature and to obtain the order parameter. Due to the strong intensity of the observed magnetic reflections, two hour scans were sufficient to determine the magnetic structure and its evolution as a function of temperature. Refinements were performed with the Rietveld method using the FullProf Suite \cite{RODRIGUEZCARVAJAL199355,Rodriguez1990}.

\section{Results}
\subsection{Structural Characterization}

\begin{figure}[h!]
\includegraphics[width=\linewidth]{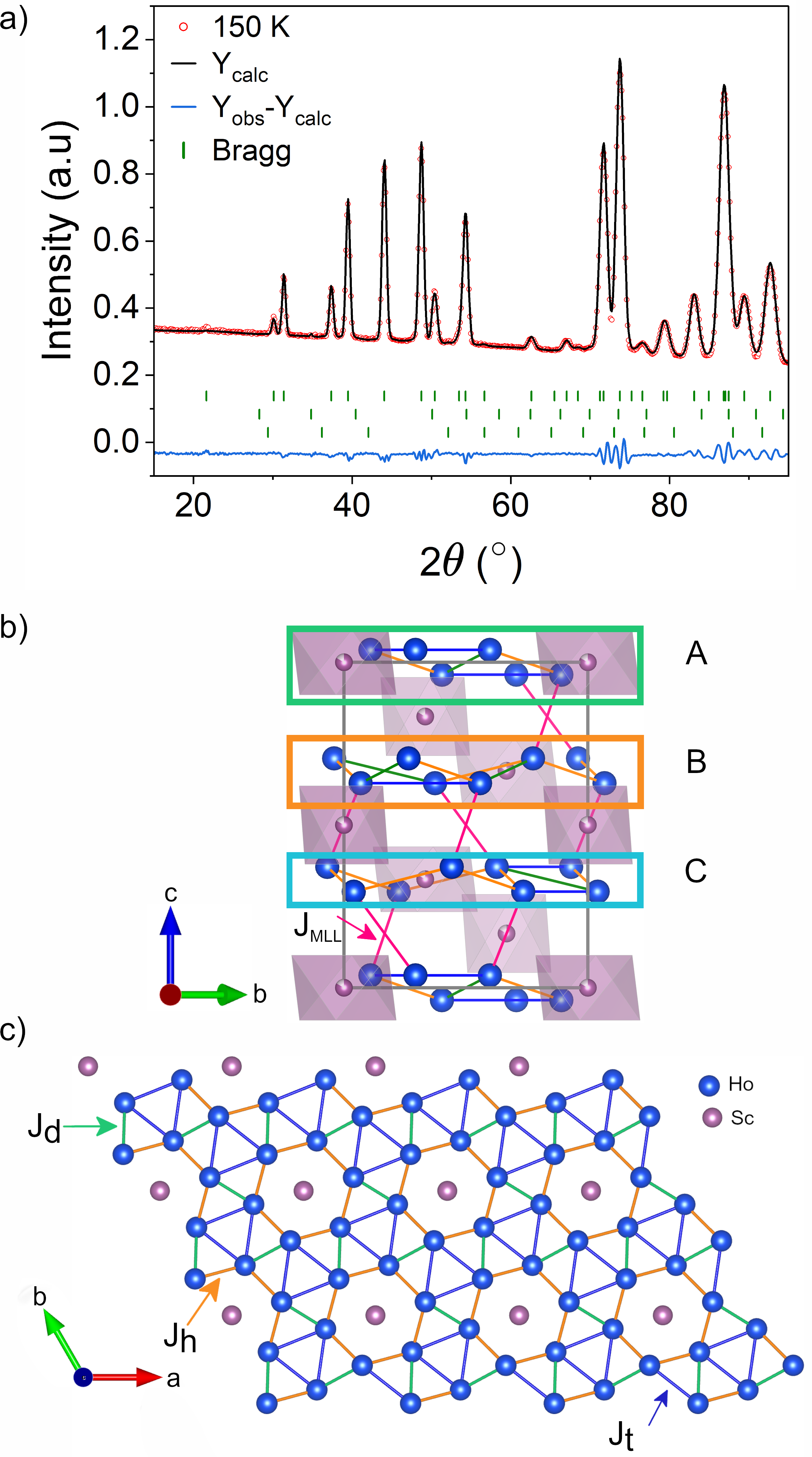} 
\captionsetup{justification=justified, singlelinecheck=false}
\caption{\justifying a) Neutron powder diffraction at 150K (red circles), Rietveld refinement (solid black line), difference between the observed and calculated pattern (solid blue line) and the Bragg positions (green vertical lines) are plotted. The first row of vertical lines represents the Bragg peak positions of the main phase (\hso) the second and third row correspond to the impurity phases of \ce{Ho2O3} and \ce{Sc2O3} respectively. b) Projection of the structure  onto the bc-plane showing Sc polyhedra (pink symbols) and Ho$^{3+}$ ions as blue spheres, three buckled MLL layers of \ce{Ho^3+} ions are arranged with an ABC-stacking along the c-axis highlighted by the green, orange and blue boxes. A possible inter-layer interaction scheme, represented by pink lines between pairs of \ce{Ho^3+} ions from different layers with a bond distance of $d_{J_{MLL}} = 3.489$ is shown. c) Projection onto the ab-plane of \hso~ showing just on MLL layer. \ce{Ho^3+} and \ce{Sc^3+} ions are represented as blue and pink spheres. The three nonequivalent nearest neighbor exchange interactions are depicted by lines with different colors whose bond distances correspond to $d_{J_{h}} = 3.497$~\AA~(orange), $d_{J_{t}}=4.062$~\AA~(blue), and  $d_{J_{D}}=3.486$~\AA~(green).}
\label{fig:str-latt}
\end{figure}

 NPD measurements confirmed that \hso~crystallizes in centrosymmetric trigonal space group (R$\bar{3}$) with no structural transition down to $T=1.6$~K. Figure \ref{fig:str-latt}(a) shows the diffractogram obtained at $T=150$~K ($\lambda=2.453$~\AA) where the cell parameters are $a=b= 9.4421(3)$~\AA, $c= 10.8957(6)$~\AA. The atomic coordinates and thermal parameters are shown in Table \ref{tab:atpos}. Refined atomic occupancies reveal that our sample has Sc deficiency with the resultant stoichiometry being Ho$_{2.99}$Sc$_{0.84}$O$_{5.90}$, furthermore the presence of impurities has been identified, representing 0.98 \%  of the sample in the form of \ce{Ho2O3} and 0.2 \% in the form of \ce{Sc2O3}. This behaviour is common for \ce{A2BB{'}O6} materials which are prone to cation non-stoichiometry \cite{BELIK201943, JM9950501995,IVANOV20062645}.

The crystal structure consists of three stacked layers (Fig. \ref{fig:str-latt}(b)) of a distorted triangular lattice of \ce{Ho^3+} ions where every 7$^{th}$ \ce{Ho^3+} is replaced by a non-magnetic \ce{Sc^3+} ion resulting in a rare example of a MLL. The \ce{Ho^3+} ions occupy one special Wyckoff position (18f) in the unit cell and they form hexagons (side length $d_{J_{H}} = 3.497$~\AA) that are connected into a quasi two-dimensional network by equilateral triangles ($d_{J_{T}}=4.062$~\AA) and dimers ($d_{J_{D}}=3.486$~\AA). Fig \ref{fig:str-latt}(c) shows a projection of one MLL layer onto the ab-plane of \hso~ where the blue and pink balls represent the \ce{Ho^3+} and \ce{Sc^3+} ions respectively, the orange lines represent the \ce{Ho^{3+}}-\ce{Ho^3+} interaction that forms the hexagons ($J_h$), blue lines represent $J_t$ forming the equilateral triangles and green lines represent the dimer interaction. These three bond distances are different in length and environment giving rise to different interaction strengths ($J_h \ne J_t \ne J_d$). Three layers of the distorted MLL are stacked along the {\bf c}-direction in each unit cell, the distance between those layers is $d_{MLL}=3.489$~\AA~ (Fig \ref{fig:str-latt}(b)) which is similar to the dimer distance $d_{J_{D}}$, this could imply a significant inter-layer interaction and thus potential 3D behaviour in the system.

\begin{table}[h]
\captionsetup{justification=justified, singlelinecheck=false}
\caption{\justifying Atomic coordinates and isotropic Debye-Waller factors obtained from the rietveld refinement of the neutron powder diffraction pattern of the powder sample at $T = 150$~K (DMC, PSI). The refinement was performed in space group $R\bar{3}$ (No. $148$) yielding lattice parameters $a=b= 9.4421(3)$~\AA, $c= 10.8957(6)$~\AA, with $R_{F}= 2.14$\% and $R_{wp}=2.23$\%.}
\label{tab:atpos}
\resizebox{\columnwidth}{!}{%
\begin{tabular}{c|c|c|c|c|c|c}
\hline
Atom  & Wyckoff &    x           &    y           &     z           &   B        &  occ.\\
\hline
\hline
Ho  & 18f &   0.399(1) & 0.106(1)  & 0.038(1) & 0.672 & 0.99(3) \\
Sc1  &  3a &   0.000  & 0.000   & 0.000  & 0.672 & 0.834(5)  \\
Sc2  &  3b &   0.000  & 0.000   & 0.500  & 0.672 & 0.834(5)  \\
O1   & 18f &   0.213(1) & 0.184(2) & 0.098(1) & 0.672 & 1.000(0)  \\
O2   & 18f &   0.185(2) & 0.159(2)  & 0.620(1) & 0.672 & 0.96(3) \\
\hline
\end{tabular}
}
\end{table}

\subsection{Magnetic Properties}

\begin{figure}[h!]
\includegraphics[width=\linewidth]{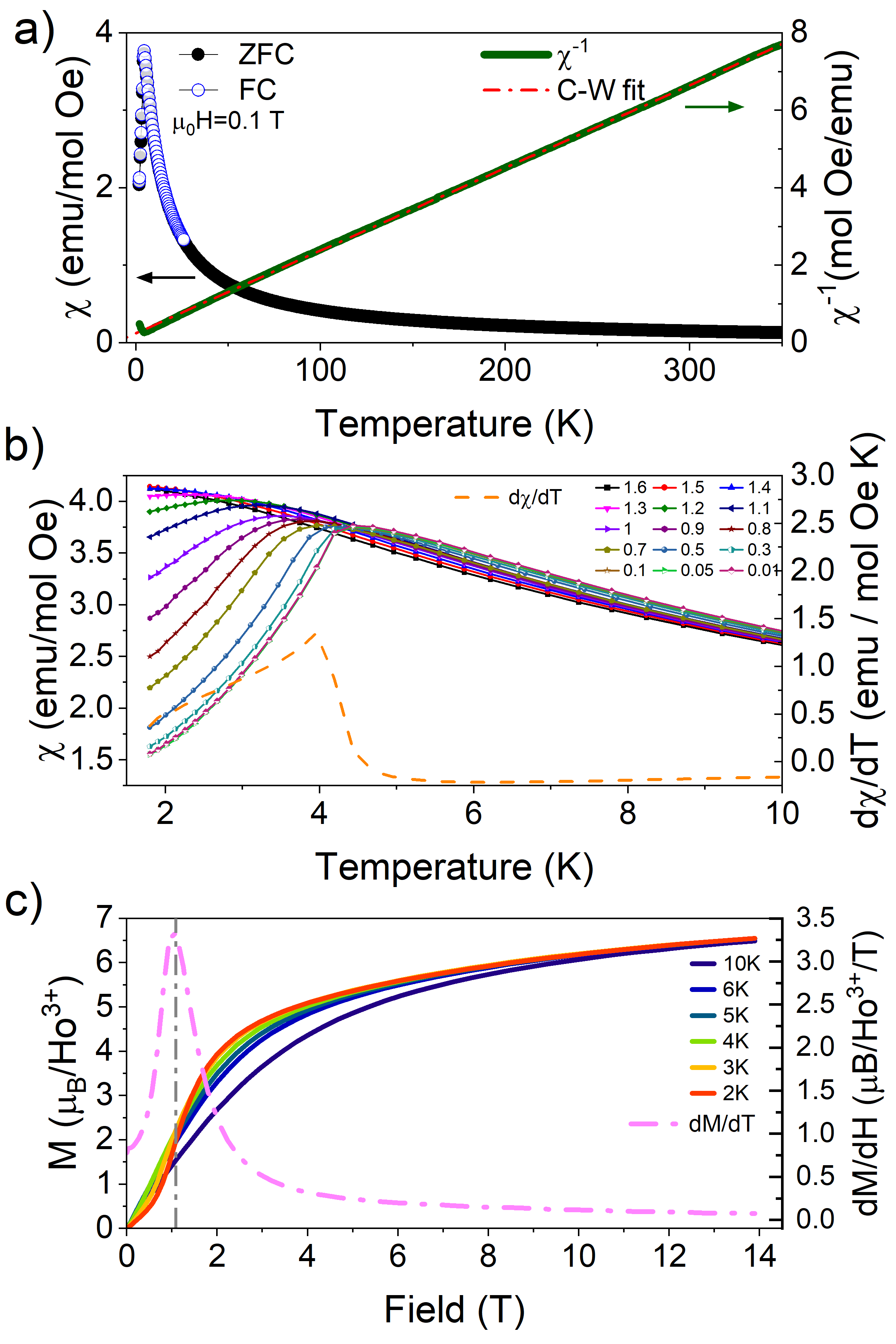}
\captionsetup{justification=justified, singlelinecheck=false}
\caption{\justifying a) DC magnetic susceptibility as a function of temperature measured with $\mu_0H=0.1$~T, the zero-field cooled (black circles) and field-cooled (blue circles) show the exact same behavior. On the right axis the inverse of the susceptibility is plotted (green line) with the Curie-Weiss law fitted (red dashed line). b) Susceptibility as a function of the temperature for various fields, the peak broadens and moves to lower temperature as the field increases until is not visible any more above 1.6~T. c) Magnetization as a function of the magnetic field at $T=2$~K shows a metamagnetic transition at $H=1.09$~T that fades in the measurement above $T=5$~K, saturation is not reached at 14~T. On the right axis the derivative of the magnetization as a function of the field at $T=2$~K is presented by the pink dashed line.} \label{fig:ChivsT}
\end{figure}

Magnetic susceptibility as a function of the temperature is presented on Fig.~\ref{fig:ChivsT}(a) where a sharp transition can be identified revealing the presence of long-range magnetic order. The temperature derivative of the susceptibility $d\chi/dT$ shows a sharp peak from which we obtain the N{\'e}el temperature of $T_N=4.0\pm0.1$~K (Fig.~\ref{fig:ChivsT}(b) left axis). The inverse of the magnetic susceptibility is also shown in Fig \ref{fig:ChivsT}(a) by the green line. It was fitted with a Curie-Weiss model at high temperatures (200-300 K) including a temperature independent constant $\chi_0$, this term stems from diamagnetic contributions from the sample.
\begin{equation}
    \chi^{-1}=\frac{T-\theta_{CW}}{\chi_{0} \cdot (T - \theta_{CW})+C}
\end{equation}
The fit yields the magnetic effective moment $\mu_{eff}=11.25\pm0.05$~$\mu_B$, a value somewhat bigger than that expected for \ce{Ho^3+} of $\mu_{eff}^{Theory}=10.4$~$\mu_B$ (based on $S=2$, $L=6$, $J=8$, $g_J=1.25$). This behavior can happen in rare earth compounds where the populations of the excited crystal field multiplets, which may have different moments compared to the ground state, increase as a function of temperature \cite{PhysRevMaterials.3.094404,BESARA201423}. The inverse of the magnetic susceptibility was also fitted to the Curie-Weiss law for the lower temperature range of 50 to 150~K now giving a value near the expected value $\mu_{eff}=10.51\pm0.03 \mu_B$. The average of $\mu_{eff}$ from the low and high temperature fits is $\mu_{eff}=10.84\pm0.04~\mu_B$.

The Curie-Weiss temperature for the low and high temperature ranges are $\theta_{CW}=-14\pm0.5$~K and $\theta_{CW}=-7\pm0.1$~K respectively, with the average value of $\theta_{CW}=-10.5\pm0.3$~K indicating the presence of antiferromagnetic interactions. The frustration parameter, defined as the absolute ratio of the Curie-Weiss temperature $\theta_{CW}$ to the ordering temperature $T_{N}$ \cite{Billington2015} is $f=10.5 /4 \sim 2.63$ illustrating the weakly frustrated nature of this system. Further measurements under different magnetic fields reveal that the N{\'e}el transition moves to a lower temperature and the peak broadens considerably with increasing field (Fig.\ref{fig:ChivsT}(b)).

Isothermal magnetization curves at $T=2,3,4,5,6$ and $10$~K with field ranging up to 14~T present no hysteresis implying an antiferromagnetic ground state. A field-induced transition is revealed by a step in the magnetization and calculated from the derivative of the magnetization to be at $H=1.09$~T (Fig.\ref{fig:ChivsT}(c)). Although the magnetization curve flattens above $\approx 3$~T, it continues to increase and does not appear to reach saturation even at 14~T, as indicated by small but finite derivative ($dM/dT$) at this field. The calculated saturation magnetization at 14~T $M_s=6.98 \pm 0.05$~$\mu_B$ is lower than the theoretical saturation value of $M=g_JJ=10$~$\mu_B$. While further increase in magnetization might be achieved with a higher field, another explanation for this result is anisotropic behavior which is common for rare earth compounds \cite{RevModPhys.82.53,SKOMSKI2009675,PhysRevB.104.214410}. The anisotropy prevents the spins from responding completely to the field in this powder sample, thus the full moment is not observed. It is worth mentioning that the number of studies investigating MLL in magnetic fields is limited \cite{PhysRevB.108.L060406}.

\subsection{Heat Capacity} 

\begin{figure}[h!]
\includegraphics[width=\linewidth]{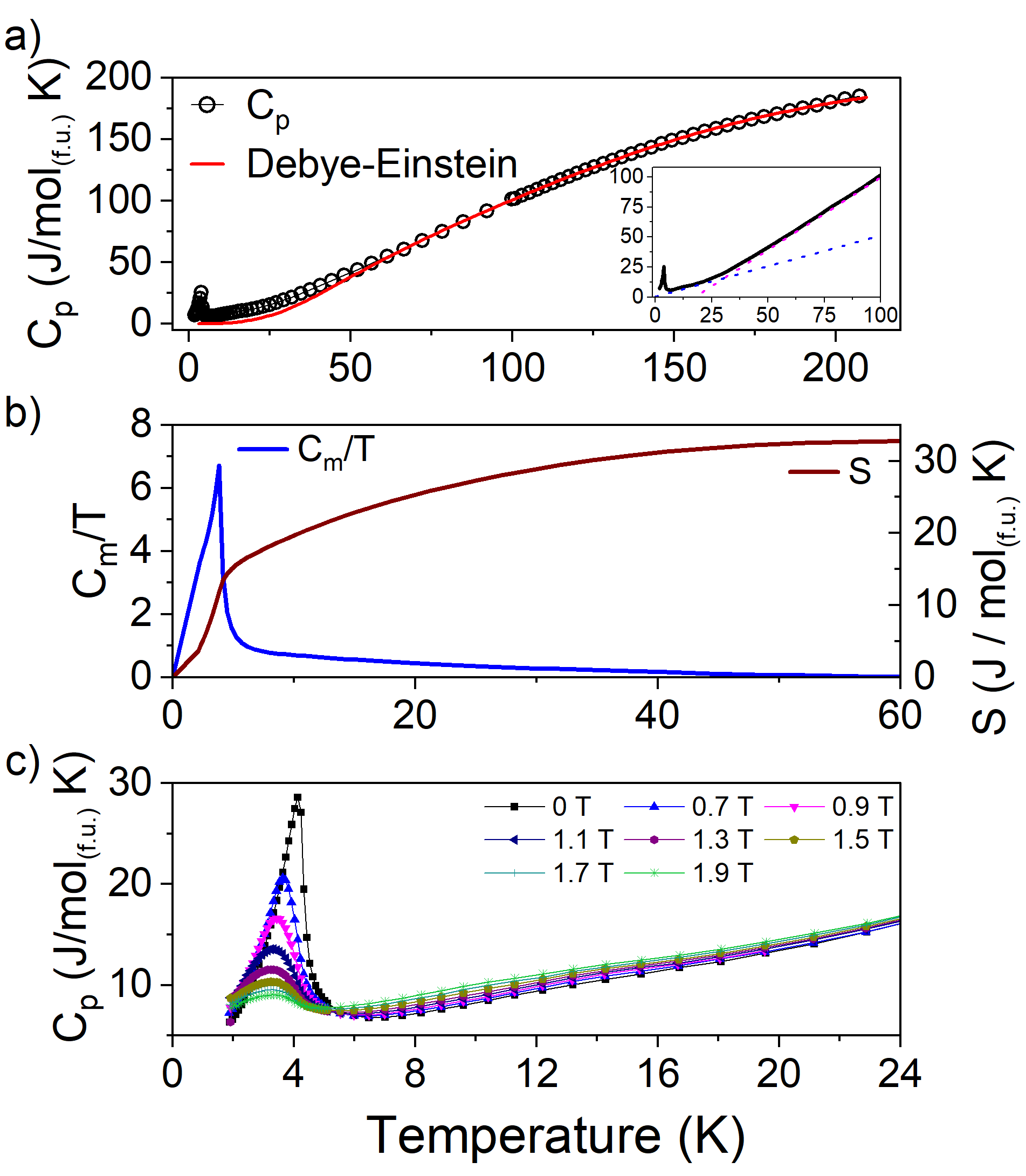}
\captionsetup{justification=justified, singlelinecheck=false}
\caption{\justifying a) Heat capacity measurement in zero field represented by black open circles. The phononic contribution is fitted by the sum of Debye and Einstein contributions, the red line represents the extrapolation of the model described in Eq~\ref{eq:2} over the whole temperature range. The inset shows the lower temperature region where a change of slope occurs at about $T=35$~K, two dotted lines are added tangential to both regions to emphasize the slope change. b) The magnetic heat capacity, obtained by subtracting the the phonon contribution, is plotted as $C_m / T$ versus $T$ (blue line). The entropy is calculated by integrating $C_m/T$ over the temperature range. It saturates to a value of $S\sim 29$~J/molK (red line). c) The low temperature heat capacity is plotted showing the evolution of the $\lambda$~anomaly into a broader peak as the field is increased. The heat capacity and entropy are displayed in units of J/molK where the mole refers to a formula unit containing three Ho$^{3+}$ ions.} \label{fig:Cp}
\end{figure}

Heat capacity measurements as a function of temperature from $T=280$~K down to $T=1.7$~K at $H=0$~T show a $\lambda$-shape peak giving at $T_N=4.1~K$ (Fig. \ref{fig:Cp}(a)) which agrees with the value obtained from susceptibility data. The phononic contributions of the heat capacity were modeled with the sum of Debye and Einstein models by the equation: 

\begin{align}\label{eq:2}
\begin{split}
&C_{lattice}=n\cdot C_{Debye}+m\cdot C_{Einstein} \\
&= 0.5\cdot 9Nk_B \left( \frac{T}{\Theta_D} \right)^3 \int_0^{\Theta_D/T} \frac{x^4 e^x}{(e^x - 1)^2} dx  \\
&+ 0.5\cdot 3Nk_B \left( \frac{\Theta_E}{T} \right)^2 \frac{e^{\Theta_E/T}}{(e^{\Theta_E/T} - 1)^2}        
\end{split}
\end{align}

where N is the number of atoms in the solid, $k_B$ is the Boltzmann constant, $\theta_D$ is the Debye temperature, $\theta_E$ is the Einstein temperature. The fit was performed over a temperature range from 100 to 200 K with $\theta_D=224$~K and $\theta_E=543$~K. The phononic contribution was extrapolated down to low temperature and subtracted from the experimental data to obtain the magnetic contribution to the heat capacity. The magnetic heat capacity divided by temperature is shown in Fig.~\ref{fig:Cp}(b) by the solid blue line. Integrated over temperature gives the magnetic entropy which is shown by the solid red line whose scale belongs to the right y-axis. The heat capacity and thus the entropy were normalized in units of J/molK where the mole refers to a formula unit containing three Ho3+ ions. The  magnetic entropy saturates at $S=32.68\pm0.05$~J/molK, a value lower than the expected value $S_{mag} = 3 R \ln(2J+1) = 70.67$~J/molK where $R$ is the gas constant, and $J=8$ for \ce{Ho^3+}. It should be noted that the factor of three reflects the fact that the formula unit contains three magnetic ions. This low value of saturation entropy is close to $3 R \ln(3)=27.4$~J/molK, suggesting a triplet in the ground state. This outcome is possible for Ho$^{3+}$ in an octahedral crystal field which would split the $J=8$ multiplet into doublets and singlets spread of $\approx 100$~meV and the ground state is often constituted by a closely spaced singlet and triplet.

Only about 46\% of the saturation entropy is recovered up to the transition temperature while full saturation is achieved at the much higher temperature of $~\approx 35$~K. This is also evident in the raw data (Fig.\ref{fig:Cp}(a) inset) with a change of slope in the region below $T=35$~K (emphasized by two tangent dotted lines tangent to each region), which can be ascribed to short-range order above the transition. Fig.\ref{fig:Cp}(c) depicts the evolution of the short-range order and the $\lambda$-anomaly with the field. The transition becomes broader and shifts to lower temperature with increasing field in agreement with the susceptibility measurements. This field-induced suppression of the transition is consistent with long-range antiferromagnetic order. In the region 6~K~$< T < 30$~K the shape of the broad peak due to short-range order is maintained but it increases as the field increases showing an enhancement of short-range fluctuations in the field.

\subsection{Magnetic Structure} 

\begin{figure}
\includegraphics[width=0.99\linewidth]{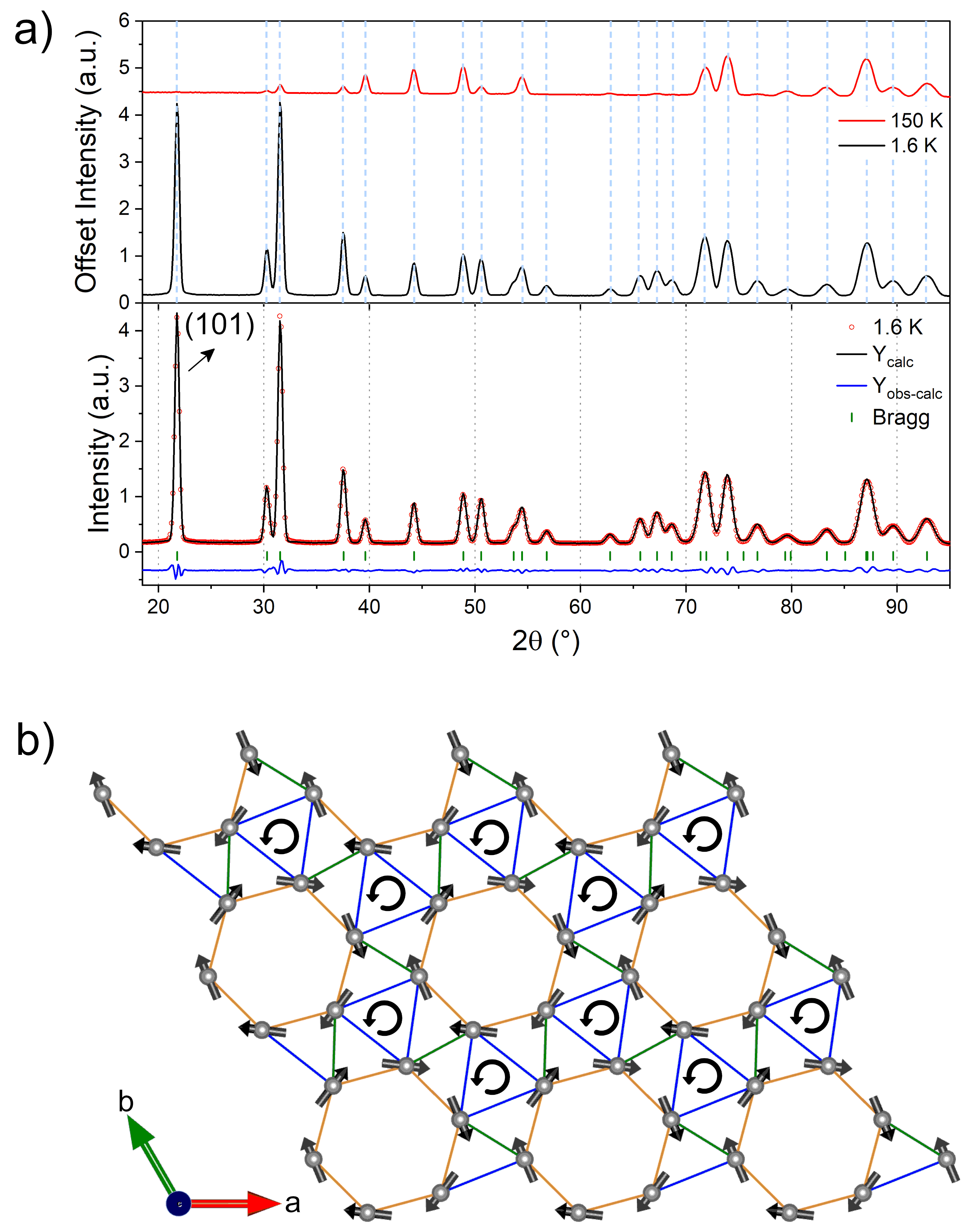}
\captionsetup{justification=justified, singlelinecheck=false}
\caption{\justifying a) Rietveld refinement of the NPD at $T=1.6$~K~$<T_N$. The data are plotted as red circles, Rietveld refinement result as a solid black line, the difference between the experimental and calculated curves is depicted as a solid blue line below the diffractogram. The vertical green lines represent the Bragg positions of the structure. The first magnetic reflection is indexed as (101). The propagation vector for the magnetic structure is $\kappa= (0 0 0)$. b) Crystal structure projected onto the ab-plane showing the Ho$^{3+}$ ions of one MLL layer. The black arrows give the spin directions of the refined magnetic structure. The nearest neighbor interactions are represent by the green ($J_d$), blue ($J_t$) and yellow ($J_h$) lines. The blue triangles have all the same positive chirality which is opposite to the chirality of the hexagons.}
\label{fig:NPD}
\end{figure}

Figure \ref{fig:NPD}(a) shows the neutron diffractogram measured below $T_N$ at $T=1.6$~K ($\lambda=2.453$~\AA). All structural peaks demonstrated an increase in intensity which can be successfully indexed by the ordering vector $\kappa=(0,0,0)$. Using this ordering vector, the magnetic representation decomposes into six one-dimensional irreducible representations repeated three times each $\Gamma_m = 3\Gamma_1 + 3 \Gamma_2 + 3\Gamma_3 +3\Gamma_4 + 3\Gamma_5 + 3\Gamma_6$. All the representations were tested and $\Gamma_2$ gave the best fitting parameters. The total magnetic moment is $M_T = 7.23~\mu_B$ which agrees with the saturation magnetization calculated previously $M_{s}=6.98~\mu_B$. As shown in Fig.~\ref{fig:NPD}(b), the resulting magnetic structure is coplanar but not colinear. It consists of antiferromagnetic spin alignment about the dimer $J_{d}$ bond (green line). The spins around the equilateral triangles coupled by $J_{t}$ form a 120$^\circ$ arrangement (blue line), they rotate in an anticlockwise sense giving a positive vector chirality on all these triangles. In contrast the spins on the hexagons ($J_{h}$, yellow line) have negative chirality and rotate clockwise.

An ideal MLL with uniform exchange interactions ($J_d=J_t=J_h$) should have the staggered vector chiral  SCV1 spin structure, where the three spins of every equilateral triangle (blue triangle) form a 120$^\circ$ structure with alternating chirality. For a non-uniform MLL (keeping the six-fold symmetry invariant) with three non-equivalent AFM exchange interactions $J_{d}$, $J_{t}$, $J_{h}$ (where $J_{d} \gg J_{t} > J_{h}$), three possible chiral spin structures PVC, NVC and SVC2 were reported depending on the relative strengths of the $J$s. \hso~ forms a non-uniform MLL containing three non-equivalent Ho$^{3+}$-Ho$^{3+}$ bonds. The bond distance $d_{J_{d}}=3.4886$~\AA~ is smallest suggesting that $J_d$ might be the strongest interaction and indeed the spins are antiparallel (180$^\circ$) around this bond. $d_{J_{t}}=3.497$~\AA~ is slightly larger and the spins form a 120$^\circ$ structure about the triangles form by this bond as expected for a strong antiferromagnetic $J_t$. Finally $d_{J_{h}}=4.062$~\AA~ is significantly larger suggesting a weak $J_h$ interaction and indeed neighboring spins around the hexagons rotate by only 60$^\circ$ suggest the spin structure does not satisfy $J_h$ probably because this interaction is weak. \hso~ is therefore compatible the $J_{d} \gg J_{t} > J_{h}$ scenario and selects the positive vector chirality (PVC) spin structure.

\begin{figure}
\includegraphics[width=0.99\linewidth]{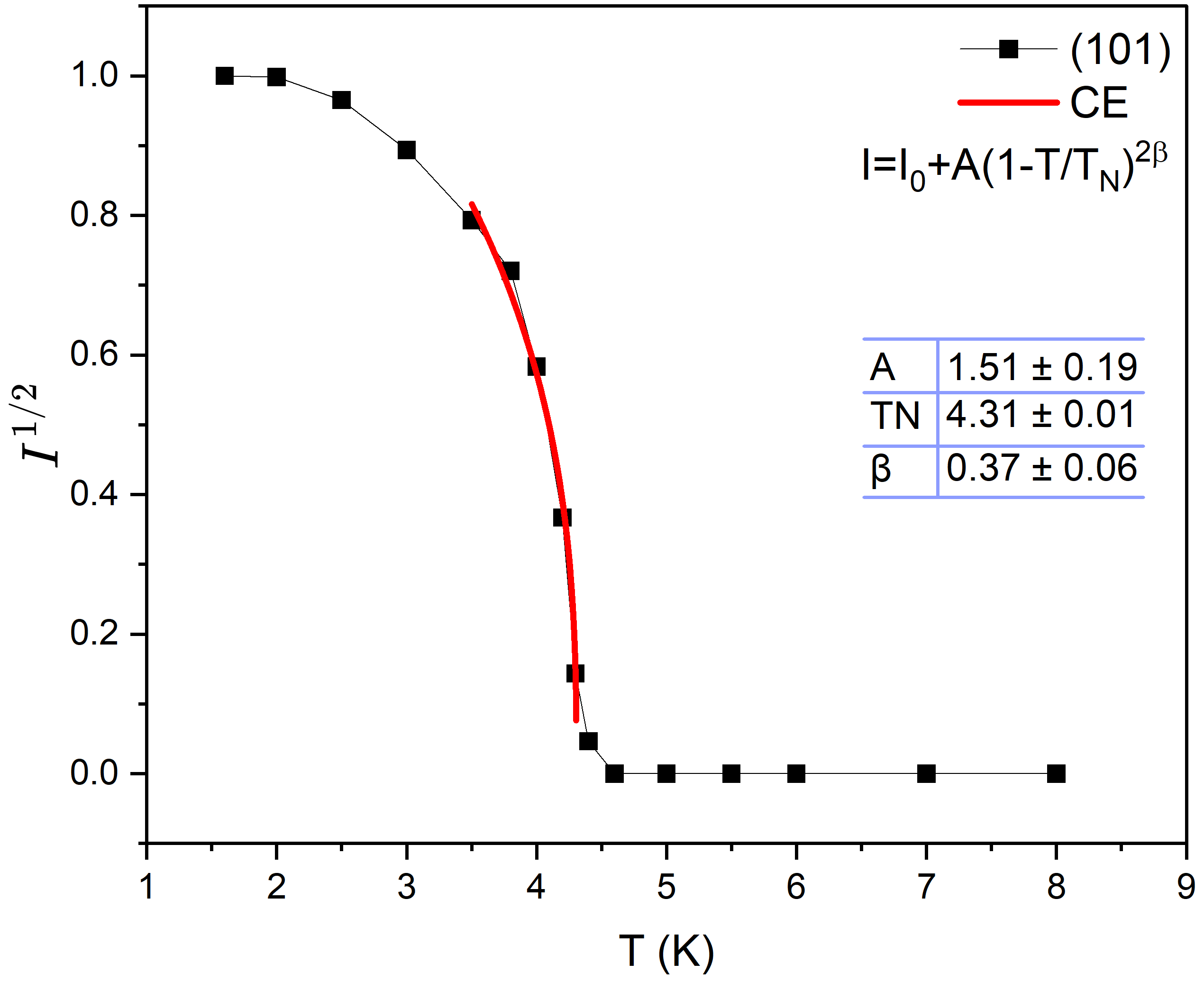}
\captionsetup{justification=justified, singlelinecheck=false}
\caption{\justifying Temperature dependence of the normalized square root of the integrated intensity of the 101 magnetic Bragg peak of \hso~is plotted as black squares. The solid red line represents the fitted curve of the critical exponent equation, the transition temperature is calculated to be $T_{N}= 4.31$~K and the critical exponent is $\beta=0.37$.}
\label{fig:IINt}
\end{figure}

Measurements at different temperatures were performed to investigate the evolution of the magnetic moment as a function of the temperature. The ordered magnetic moment below the magnetic phase transition at $T<T_{N}$ is given by $m\sim \sqrt(I)$ where $I$ is the intensity of a magnetic reflection. The square root of the integrated intensity of the (101) magnetic reflection of \hso~ is normalized and plotted as a function of temperature in Fig.~\ref{fig:IINt}. The order parameter (OP) varies as a continuous function of $T$ indicating a second order magnetic phase transition, the OP has non-zero values in the magnetically long-range-ordered state and reduces as it approaches the paramagnetic state. Magnetic systems are classified into several categories with different values of temperature exponent $\beta$ in $m\sim (T_N-T)^{\beta}$, and depending on the value of this exponent it is possible to determine the magnetic properties of \hso. For 2D Ising behaviour, $\beta=1/8$ is expected, while for a 3D Ising $\beta=0.326$, whereas for a 3D Heisenberg system the expected exponent would be $\beta=0.367$ and for mean-field behaviour $\beta=0.50$. The equation was fitted over the temperature range from 3.5 to 4.4 K, yielding to a value of $T_{N}=4.31$~K and $\beta=0.37$. Taking into account the uncertainty, it is not possible to distinguish between Ising and Heisenberg behavior. However, it can be concluded that this material exhibits 3D behavior, indicating that the interaction between layers ($J_{MLL}$) plays a significant role in the system.

\subsection{Summary}

A powder sample of \hso~was synthesized by solid state reaction. It crystallizes in a corundum related structure with a centrosymmetric trigonal space group (R$\bar{3}$) and cell parameters a=b=9.4421(3)~\AA, c=10.8957(6)~\AA. The \ce{Ho^3+} ions occupy one special Wyckoff position (18f) and  form hexagons that are connected into a quasi-2D network by equilateral triangles to form a distorted MLL. In zero magnetic field long range magnetic order is evident below $T_N=4.5$~K. Three non-equivalent \ce{Ho^{3+}-Ho^3+} magnetic exchange interactions $J_{d}$, $J_{t}$, and $J_{h}$ can be identified where $J_{d}$ has shortest \ce{Ho^{3+}-Ho^3+} distance $d_{J_{d}} = 3.486$~\AA, $J_{t}$ has a slightly longer distance $d_{J_{t}} =3.497$~\AA, and $J_{h}$ has the longest distance with $d_{J_{h}} =4.062$~\AA. The ground state expected for a MLL with $J_{d}=J_{t}=J_{h}$ is a coplanar arrangement with the staggered vector chirality 1 (SVC1) structure. Haraguchi et al. \cite{PhysRevB.98.064412} proposed three different ground states for a MLL with $J_{d} \gg J_{t} > J_{h}$ namely SVC2 with staggered vector chirality, PVC with uniform positive vector chirality while NVC has uniform negative vector chirality. From the Rietveld refinement of the NPD measurements, the magnetic structure of \hso~ is depicted in Fig.~\ref{fig:NPD} where $J_{d}$ and $J_{t}$ are strongly antiferromagnetic coupling antiparallel spin pairs and 120$^\circ$ order around the equilateral triangles respectively, while $J_{h}$ appears weak and does not influence the structure. The structure thus realizes a dimer phase where antiferromagnetic dimers are coupled together by frustrated triangular interactions \cite{PhysRevB.109.184422,PhysRevB.105.L180412}. Following the formula for the vector chirality mentioned by Okuma et al. \cite{PhysRevB.95.094427} the spins of \hso~ order in the positive vector chirality (PVC) structure. The extracted value of the critical exponent $\beta$ indicates that this system exhibits three-dimensional behaviour thereby implying a significant $J_{MLL}$ interaction. Further investigations of \hso~could focus on the crystal field levels and understanding of the ground state wavefunction of the \ce{Ho^3+} ion. The effect of magnetic field on the structure may also be investigated, since a rich series of field induced-phases is predicted \cite{PhysRevB.108.L060406}.

\section{Acknowledgements}
We acknowledge the CoreLab Quantum Materials, Helmholtz Zentrum f{\"u}r Materialen und Energie (HZB), Germany, where the powder sample was synthesized and its magnetic and thermodynamic properties were measured. We also acknowledge the Swiss spallation neutron source (SINQ) at the Paul Scherrer Institute (PSI), Switzerland. This research is supported by the Deutsche Forschungsgemeinschaft (DFG) through the project B06 of the SFB 1143 (ID 247310070).

\bibliographystyle{apsrev4-2}
\bibliography{bib}

\end{document}